\documentclass[aps,prd,eqsecnum,preprint,tightenlines,nofootinbib,showpacs]{revtex4-1}
\usepackage{graphicx,amsmath,latexsym}

\def\bea{\begin{eqnarray}}
\def\eea{\end{eqnarray}}
\def\be{\begin{equation}}
\def\ee{\end{equation}}
\newcommand{\ub}[1]{\underline{#1}}
\newcommand{\Pminus}{{\cal P}^-}

\begin{document}

\title{A first nonperturbative calculation in light-front QED \\
for an arbitrary covariant gauge}

\author{Sophia S. Chabysheva}
\author{John R. Hiller}
\affiliation{Department of Physics \\
University of Minnesota-Duluth \\
Duluth, Minnesota 55812}

\date{\today}

\begin{abstract}
This work is the first check of gauge invariance for nonperturbative
calculations in light-front QED.
To quantize QED in an arbitrary covariant gauge, we use a light-front 
analog of the equal-time Stueckelberg quantization.  Combined with
a Pauli--Villars regularization, where massive, negative-metric
photons and fermions are included in the Lagrangian, we are then able
to construct the light-front QED Hamiltonian and the associated
mass eigenvalue problem in a Fock-space representation.  The formalism
is applied to the dressed-electron state, with a Fock-space truncation
to include at most one photon.  From this eigenstate, we compute
the anomalous magnetic moment.  The result is found to be
gauge independent, to an order in $\alpha$ consistent with the
truncation.
\end{abstract}

%%%%%%%%%%%%%%%%%%%%%%%%%%%%%%%%%%%%%%%%%%%%%%%%%%%%%
% 10. THE PHYSICS OF ELEMENTARY PARTICLES AND FIELDS
% 11.10.-z Field theory (for gauge field theories, see 11.15)
% 11.10.Ef Lagrangian and Hamiltonian approach
% 11.10.Gh Renormalization
% 11.15.-q Gauge field theories
% 11.15.Tk Other nonperturbative techniques
% 12. Specific theories and interaction models; particle systematics
% 12.38.-t Quantum chromodynamics see also 24.85 Quarks, gluons,
% and QCD in nuclei and nuclear processes
% 12.38.Lg Other nonperturbative calculations
%%%%%%%%%%%%%%%%%%%%%%%%%%%%%%%%%%%%%%%%%%%%%%%%%%%%%%%

%

\maketitle

%%%%%%%%%%%%%%%%%%%%%%
\section{Introduction}
\label{sec:Introduction}
%%%%%%%%%%%%%%%%%%%%%%

Any calculation in a gauge theory should be checked for its
gauge dependence.  Unfortunately, nonperturbative calculations
in light-front QED~\cite{DLCQreview} have been limited to a 
single gauge, usually light-cone gauge.  This is due to the 
need to solve the constraint equation for the nondynamical
part of the fermion field, which is entangled with the
photon field.  A careful use of Pauli--Villars (PV)
regularization~\cite{PauliVillars} has been shown~\cite{OnePhotonQED}
to allow the use of Feynman gauge, by providing cancellation
of the photon-field dependence in the constraint equation.
What is remarkable, however, is that this cancellation is
actually not unique to Feynman gauge but holds for any
gauge.  Thus, nonperturbative calculations can be done in any
gauge, provided the free-photon part of the Hamiltonian can
be constructed.  Here we provide such a construction for an
arbitrary covariant gauge and apply the formalism to a
calculation of the dressed-electron eigenstate and its
anomalous magnetic moment, in order to investigate the gauge
invariance of the result.

This builds on earlier work on Yukawa theory~\cite{bhm}
and QED~\cite{OnePhotonQED,ChiralLimit,Thesis,SecDep,TwoPhotonQED,VacPol},
where PV particles are used to regulate a light-front 
Hamiltonian and the eigenstates of the Hamiltonian are
computed in one or more charge sectors of the theory.
The eigenstate is expanded in a truncated Fock basis.  The
eigenvalue problem becomes a coupled set of integral
equations for the wave functions, which are the coefficients
of the Fock states in the expansion.  Truncation keeps the
coupled set finite in size, and the PV regularization keeps
the integrations finite.  For severe truncations, the
coupled set can be solved analytically.  In general, the
set is solved numerically~\cite{Thesis,TwoPhotonQED}.  The
renormalization can be handled in a standard way, with
the bare parameters of the original Lagrangian fixed
by physical constraints, or by a sector-dependent
parameterization, where the bare parameters become
dependent on the Fock sectors connected by the
terms in the Hamiltonian~\cite{Wilson,hb,Karmanov,Karmanov2,Vary,SecDep}.
However, in a weakly coupled theory such as QED,
such an approach cannot be expected to compete with
high-order perturbation theory, due to numerical errors.
Consideration of QED is a test for a method intended
for strongly coupled theories.

Light-cone coordinates~\cite{Dirac,DLCQreview} are used
in order to have well-defined Fock-state expansions,
which are at the heart of the method.  We define 
these coordinates as $x^+=t+z$ for time and 
$\ub{x}=(x^-=t-z,\vec x_\perp)$ for space, with
$\vec x_\perp=(x,y)$.  The light-cone energy
is $p^-=E-p_z$ and momentum, $\ub{p}=(p^+=E+p_z,\vec p_\perp)$.
The mass-shell condition $p^2=m^2$ relates these as
$p^-=(m^2+p_\perp^2)/p^+$.  The positivity of $p^+$
keeps the vacuum simple and prevents vacuum contributions
to the Fock expansions, except for the possibility
of zero modes~\cite{ZeroModes}.  These modes of
zero $p^+$ can be neglected in theories where
symmetry breaking does not occur.

Our new construction of the Hamiltonian is based on
a light-front analog of the equal-time Stueckelberg
quantization of a massive vector 
field~\cite{Stueckelberg}.\footnote{For an
alternative construction for the massless case,
which uses the canonical Dirac constraint procedure,
see \protect\cite{Srivastava}.}  The Stueckelberg
quantization is known to allow for a zero-mass limit.
It is also useful as a way to treat the physical
and PV photons on an equal footing, consistent with
the need to maintain the PV regularization.  The 
quantization adds a fourth (unphysical) polarization
to the three physical polarizations. The unphysical
polarization is the only one that does not satisfy
the Lorentz gauge condition $\partial\cdot A=0$.  However,
it does satisfy the Euler--Lagrange field equation,
because its four-momentum is placed on a different,
gauge-dependent mass shell, chosen in just such a
way as to satisfy the field equation.  The key
to the light-front analog is that the chosen
mass shell is invoked for the minus component
of the momentum, rather than the zero component.
The details of this can be found below, in Sec.~\ref{sec:LightFrontQED}.

With this new quantization, we can formulate mass
eigenvalue problems for the eigenstates of QED in an
arbitrary covariant gauge and test for the gauge
invariance of physical quantities computed from
the eigenstates.  We expect that gauge invariance
will be broken by approximations made in solving the
eigenproblems.  One such approximation is the
Fock-space truncation used to reduce the eigenproblem
to a finite size.  Another is retention of finite
values for the regulating PV masses; the regularization
is constructed with use of flavor-changing currents
that explicitly break gauge invariance~\cite{OnePhotonQED}
and are removed only in the infinite-PV-mass limit, 
which may not be possible in a numerical calculation.

As a first test, we apply this formalism to a calculation
of the dressed-electron eigenstate.  Fock space is
truncated to include only the bare-electron state
and the one-electron/one-photon states, plus their
PV analogs.  This leads to an analytically solvable
problem, reduced to an effective $2\times2$ matrix 
problem in the one-electron sector.  From the solution,
the anomalous moment can be computed, from the
zero-momentum-transfer limit of the spin-flip
transition amplitude.

To obtain meaningful results, it is important to
maintain the chiral symmetry of the 
massless-electron limit.  This is achieved by
adjusting the coupling strengths of the PV photons,
to ensure that the dressed mass is zero when the
bare mass is zero.  As will be seen below, in 
Sec.~\ref{sec:DressedElectron}, this requires
two PV-photon flavors in any gauge.  However,
one flavor is sufficient if the limit of infinite PV-fermion
mass is taken. This also holds for the
sector-dependent approach~\cite{Karmanov2}, as we show in
Sec.~\ref{sec:DressedElectron}.  In general,
the constraint of chiral-symmetry restoration
and the PV-photon couplings are gauge dependent.

There are, of course, other nonperturbative methods.
Lattice gauge theory~\cite{Lattice} is particularly successful, and
use of Dyson--Schwinger equations~\cite{DSE} has produced notable results.
However, these lack the direct access to wave functions in Minkowski
space, which light-front Hamiltonian methods provide~\cite{DLCQreview}.
Thus, the methods are quite complementary, particularly now
that light-front calculations can be done in an arbitrary gauge.
There are also a light-front lattice method, the transverse
lattice~\cite{TransLattice}; a light-front approach in terms
of effective fields~\cite{Glazek}; and a supersymmetric
formulation for discrete light-front Hamiltonians specifically
for supersymmetric theories~\cite{SDLCQ}.

The remainder of the paper contains the following sections.
The general formalism for an arbitrary covariant gauge is
given in Sec.~\ref{sec:LightFrontQED}.  The dressed-electron
eigenproblem is solved in Sec.~\ref{sec:DressedElectron} and
used there to compute the anomalous moment.  Section~\ref{sec:Summary}
provides a summary of the method and of the results obtained.
Some details are left to three Appendices.

%%%%%%%%%%%%%%%%%%%%%%%%%%%%%%%%%%%%%%%%%%%%%%%%%%%%%%%%%
\section{Light-front QED in an arbitrary covariant gauge}
\label{sec:LightFrontQED}
%%%%%%%%%%%%%%%%%%%%%%%%%%%%%%%%%%%%%%%%%%%%%%%%%%%%%%%%%%

We begin with the QED Lagrangian for Lorentz gauge with an
arbitrary gauge parameter $\zeta$ and additional PV fields:
\bea  \label{eq:Lagrangian}
{\cal L} &=&  \sum_{i=0}^2 (-1)^i \left[-\frac14 F_i^{\mu \nu} F_{i,\mu \nu} 
         +\frac12 \mu_i^2 A_i^\mu A_{i\mu} 
         -\frac12 \zeta \left(\partial^\mu A_{i\mu}\right)^2\right] \\
&& + \sum_{i=0}^2 (-1)^i \bar{\psi_i} (i \gamma^\mu \partial_\mu - m_i) \psi_i 
  - e \bar{\psi}\gamma^\mu \psi A_\mu . \nonumber
\eea
Here
\be \label{eq:NullFields}
  \psi =  \sum_{i=0}^2 \sqrt{\beta_i}\psi_i, \;\;
  A_\mu  = \sum_{i=0}^2 \sqrt{\xi_i}A_{i\mu}, \;\;
  F_{i\mu \nu} = \partial_\mu A_{i\nu}-\partial_\nu A_{i\mu} ,
\ee
$i=0$ corresponds to a physical field, and $i=1$ and 2, to
PV fields.  The photon fields have mass $\mu_i$, and the
zero-mass limit $\mu_0\rightarrow0$ for the physical photon
is to be taken later.  

The coupling coefficients $\beta_i$
and $\xi_i$ satisfy constraints.  To keep $e$ as the charge
of physical fermion, we set $\beta_0=1$ and $\xi_0=1$.  To
regulate ultraviolet divergences that come from loop integrals,
we arrange cancellations for each internal line summed over
physical and PV fields, by imposing the constraints
\be
\sum_{i=0}^2(-1)^i\xi_i=0, \;\;
\sum_{i=0}^2(-1)^i\beta_i=0. \;\;
\ee
Two remaining coefficients, say $\xi_2$ and $\beta_2$, are fixed
by requiring chiral-symmetry restoration in the 
massless-electron limit~\cite{ChiralLimit}
and a zero photon eigenmass~\cite{VacPol}.

The dynamical fields are $\psi_{i+}$ and $A_{i\mu}$.  We
quantize the dynamical fermion fields in the usual way
\be
\psi_{i+}=\frac{1}{\sqrt{16\pi^3}}\sum_s\int d\ub{k} \chi_s
  \left[b_{is}(\ub{k})e^{-i\ub{k}\cdot\ub{x}}
        +d_{i,-s}^\dagger(\ub{k})e^{i\ub{k}\cdot\ub{x}}\right],
\ee
with
\bea
\{b_{is}(\ub{k}),b_{i's'}^\dagger(\ub{k}'\}
   &=&(-1)^i\delta_{ii'}\delta_{ss'}\delta(\ub{k}-\ub{k}'), \\
\{d_{is}(\ub{k}),d_{i's'}^\dagger(\ub{k}'\}
   &=&(-1)^i\delta_{ii'}\delta_{ss'}\delta(\ub{k}-\ub{k}').
\eea
For the vector fields, we apply a light-front analog of
Stueckelberg quantization~\cite{Stueckelberg}.

Consider the Lagrangian of a free massive vector field:
\be
{\cal L}=-\frac14 F^2+\frac12\mu A^2-\frac12\zeta(\partial\cdot A)^2.
\ee
The Euler--Lagrange field equation is
\be
(\Box +\mu^2)A_\mu-(1-\zeta)\partial_\mu(\partial\cdot A)=0.
\ee
This equation is satisfied by the Fourier expansion
\bea
A_\mu(x)&=&\int\frac{d\ub{k}}{\sqrt{16\pi^3 k^+}}\left\{\sum_{\lambda=1}^3
   e_\mu^{(\lambda)}(\ub{k})\left[ a_\lambda(\ub{k})e^{-ik\cdot x}
            + a_\lambda^\dagger(\ub{k})e^{ik\cdot x}\right]\right. \\
&& \left.+e_\mu^{(0)}(\ub{k})\left[ a_0(\ub{k})e^{-i\tilde k\cdot x}
            + a_0^\dagger(\ub{k})e^{i\tilde k\cdot x}\right]\right\}, \nonumber
\eea
with $\tilde k$ a four-vector associated with a different mass
$\tilde\mu\equiv\mu/\sqrt{\zeta}$, such that
\be
\ub{\tilde k}=\ub{k}, \;\; \tilde k^-=(k_\perp^2+\tilde\mu^2)/k^+ .
\ee
The polarization vectors are defined by
\bea
e^{(1,2)}(\ub{k})&=&(0,2 \hat e_{1,2}\cdot \vec{k}_\perp/k^+,\hat e_{1,2}), \\
e^{(3)}(\ub{k})&=&((k_\perp^2-\mu^2)/k^+,k^+,\vec k_\perp)/\mu, \\
e^{(0)}(\ub{k})&=&\tilde k/\mu, 
\eea
and satisfy $k\cdot e^{(\lambda)}=0$ 
and $e^{(\lambda)}\cdot e^{(\lambda')}=-\delta_{\lambda\lambda'}$
for $\lambda,\lambda'=1,2,3$.
The first term in $A_\mu$ satisfies $(\Box +\mu^2)A_\mu=0$ and
$\partial\cdot A=0$ separately.  The $\lambda=0$ term violates each, 
but the field equation is satisfied.  The gauge condition $\partial\cdot A=0$
is to be satisfied by projection of states onto a physical subspace.

The light-front Hamiltonian density is
\be
{\cal H}={\cal H}|_{\zeta=1}
  +\frac12(1-\zeta)(\partial\cdot A)(\partial\cdot A
                                -2\partial_-A^- -2\partial_\perp\cdot\vec A_\perp),
\ee
with the Feynman-gauge piece being
\be
{\cal H}|_{\zeta=1}=\frac12\sum_{\mu=0}^3 \epsilon^\mu
       \left[(\partial_\perp A^\mu)^2+\mu^2 (A^\mu)^2\right].
\ee
The metric of each field component is defined by
$\epsilon^\mu=(-1,1,1,1)$.
The light-front Hamiltonian for the free massive field is then found to be
\be
 \Pminus=\int d\ub{x}{\cal H}|_{x^+=0}
  =\int d\ub{k} \sum_\lambda \epsilon^\lambda \frac{k_\perp^2+\mu_\lambda^2}{k^+}
                a_\lambda^\dagger(\ub{k})a_\lambda(\ub{k}),
\ee
with $\mu_\lambda=\mu$ for $\lambda=1,2,3$, 
but $\mu_0=\tilde\mu=\mu/\sqrt{\zeta}$.
The nonzero commutators are
\be
[a_\lambda(\ub{k}),a_{\lambda'}^\dagger(\ub{k}')]
     =\epsilon^\lambda \delta_{\lambda\lambda'}\delta(\ub{k}-\ub{k}').
\ee
Thus, the Hamiltonian for the free photon field takes the usual
form except that the mass of the fourth polarization is
different and gauge dependent and that the metric of this
polarization is opposite that of the other polarizations.
In Feynman gauge, this reduces to the usual Gupta--Bleuler
quantization~\cite{GuptaBleuler}.

The nondynamical components of the fermion fields satisfy the
constraints ($i=0,1,2$)
\bea \label{eq:FermionConstraint}
i(-1)^i\partial_-\psi_{i-}&+&e A_-\sqrt{\beta_i}\sum_j\psi_{j-}  \\
  &=&(i\gamma^0\gamma^\perp)
     \left[(-1)^i\partial_\perp \psi_{i+}-ie A_\perp\sqrt{\beta_i}\sum_j\psi_{j+}\right] 
      -(-1)^i m_i \gamma^0\psi_{i+} . \nonumber
\eea
Ordinarily, light-cone gauge ($A_-=0$) is chosen, to make the
constraint explicitly invertible.  However, the interaction Lagrangian has
been arranged in just such a way that the $A$-dependent terms
can be canceled between the three constraints~\cite{OnePhotonQED}.
Multiplication by $(-1)^i\sqrt{\beta_i}$ and a sum over $i$ yields
\be
i\partial_-\psi_-
  =(i\gamma^0\gamma^\perp)
     \partial_\perp \psi_+
      - \gamma^0\sum_i\sqrt{\beta_i}m_i\psi_{i+},
\ee
as the constraint for the composite field that appears
in the interaction Lagrangian.  This constraint is the
same as the free-fermion constraint, in any gauge, and
the interaction Hamiltonian can be constructed from the
free-field solution.

\begin{figure}[ht]
\vspace{0.2in}
\centerline{\includegraphics[width=13cm]{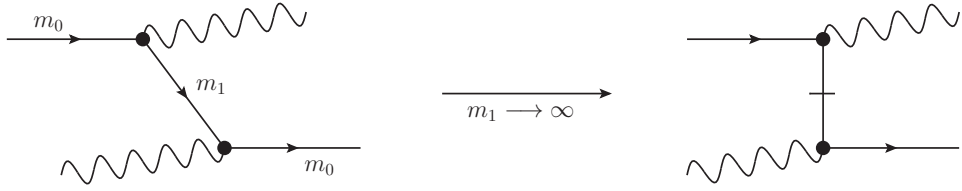}}
\caption{\label{fig:Fig1} 
The infinite-PV-mass limit of a tree graph involving
an intermediate PV fermion of mass $m_1$.
The external fermions are the physical ones, with mass $m_0$.
The limit yields an instantaneous interaction, denoted 
by the bar through the intermediate line.}
\end{figure}

Without this cancellation of $A$-dependent terms, the 
constraint would generate four-point interactions
between fermion and photon fields, the 
instantaneous-fermion interactions~\cite{DLCQreview}.
The addition of the PV-fermion fields has, in effect,
factorized these interactions into flavor-changing
photon emission and absorption three-point vertices.
The instantaneous interactions are recovered in the
limit of infinite PV fermion masses, because the light-cone
energy denominator with an intermediate PV fermion cancels the PV-mass
factors in the emission and absorption vertices,
as illustrated in Fig.~\ref{fig:Fig1}.
%%%%%%%%%%%%%%%%%%%%%%%%%%%%%%%%%%%%%
%%%%%%%%%%%%%%%%%%%%%%%%%%%%%%%%%%%%%%%%%%%%%%%%%%%%%%%%%%%%%%

The light-front Hamiltonian, without antifermion terms, is
\bea \label{eq:QEDP-}
\lefteqn{\Pminus=
   \sum_{is}\int d\ub{p}
      \frac{m_i^2+p_\perp^2}{p^+}(-1)^i
          b_{is}^\dagger(\ub{p}) b_{is}(\ub{p})} \\
   && +\sum_{l\lambda}\int d\ub{k}
          \frac{\mu_{l\lambda}^2+k_\perp^2}{k^+}(-1)^l\epsilon^\lambda
             a_{l\lambda}^\dagger(\ub{k}) a_{l\lambda}(\ub{k})
          \nonumber \\
   && +\sum_{ijls\lambda}\sqrt{\beta_i\beta_j\xi_l}\int d\ub{p} d\ub{q}\left\{
      b_{is}^\dagger(\ub{p}) \left[ b_{js}(\ub{q})
       V^\mu_{ij,2s}(\ub{p},\ub{q})\right.\right.\nonumber \\
      &&\left.\left.\rule{0.5in}{0in}
+ b_{j,-s}(\ub{q})
      U^\mu_{ij,-2s}(\ub{p},\ub{q})\right] \epsilon_{l\mu}^{(\lambda)}(\ub{q}-\ub{p})
            a_{l\lambda}^\dagger(\ub{q}-\ub{p})
                    + H.c.\right\} . \nonumber
\eea
The instantaneous-photon terms associated with light-cone gauge
do not appear.  
The polarization vectors $e_l^{(\lambda)}$ have an additional flavor
index $l$, because they depend on the mass of the photon flavor.
The vertex functions are given by~\cite{OnePhotonQED}
\bea \label{eq:vertices}
    V^0_{ij\pm}(\ub{p},\ub{q}) &=& \frac{e}{\sqrt{16 \pi^3 }}
                   \frac{ \vec{p}_\perp\cdot\vec{q}_\perp
                      \pm i\vec{p}_\perp\times\vec{q}_\perp
                       + m_i m_j + p^+q^+}{p^+q^+\sqrt{q^+-p^+}} , \\
    V^3_{ij\pm}(\ub{p},\ub{q}) &=& \frac{-e}{\sqrt{16 \pi^3}}
                        \frac{ \vec{p}_\perp\cdot\vec{q}_\perp
                      \pm i\vec{p}_\perp\times\vec{q}_\perp
                       + m_i m_j - p^+q^+ }{p^+q^+\sqrt{q^+-p^+}} , \nonumber\\
    V^1_{ij\pm}(\ub{p},\ub{q}) &=& \frac{e}{\sqrt{16 \pi^3}}
       \frac{ p^+(q^1\pm i q^2)+q^+(p^1\mp ip^2)}{p^+q^+\sqrt{q^+-p^+}} , \nonumber\\
    V^2_{ij\pm}(\ub{p},\ub{q}) &=& \frac{e}{\sqrt{16 \pi^3}}
      \frac{ p^+(q^2\mp i q^1)+q^+(p^2\pm ip^1)}{p^+q^+\sqrt{q^+-p^+}} , \nonumber
\eea
and
\bea \label{eq:vertices2}
    U^0_{ij\pm}(\ub{p},\ub{q}) &=& \frac{\mp e}{\sqrt{16 \pi^3}}
       \frac{m_j(p^1\pm ip^2)-m_i(q^1\pm iq^2)}{p^+q^+\sqrt{q^+-p^+}} , \\
    U^3_{ij\pm}(\ub{p},\ub{q}) &=& \frac{\pm e}{\sqrt{16 \pi^3}}
       \frac{m_j(p^1\pm ip^2)-m_i(q^1\pm iq^2)}{p^+q^+\sqrt{q^+-p^+}} , \nonumber\\
    U^1_{ij\pm}(\ub{p},\ub{q}) &=& \frac{\pm e}{\sqrt{16 \pi^3}}
                            \frac{m_iq^+-m_jp^+ }{p^+q^+\sqrt{q^+-p^+}} , \nonumber\\
    U^2_{ij\pm}(\ub{p},\ub{q}) &=& \frac{i e}{\sqrt{16 \pi^3}}
                     \frac{m_iq^+-m_jp^+ }{p^+q^+\sqrt{q^+-p^+}} . \nonumber
\eea
The extension to include antifermion terms is straightforward~\cite{VacPol}.
We now apply this formalism to a nonperturbative calculation of the 
dressed-electron state and its anomalous magnetic moment.

%%%%%%%%%%%%%%%%%%%%%%%%%%%%%%%%%%%%%%%%%
\section{Dressed-electron eigenstate}
\label{sec:DressedElectron}
%%%%%%%%%%%%%%%%%%%%%%%%%%%%%%%%%%%%%%%%%

\subsection{Eigenvalue problem} \label{sec:EigenvalueProblem}

We wish to solve the light-front eigenvalue problem 
$\Pminus|\psi(\ub{P})\rangle=\frac{M^2}{P^+}|\psi(\ub{P})\rangle$.
The eigenstate $|\psi(\ub{P})\rangle$ is expanded in a 
Fock basis where $P^+$ is diagonal and $\vec P_\perp$ is zero.
Here we consider the lowest order truncation, to include
the bare-electron state and the one-electron/one-photon states.
For a total $J_z=\pm\frac12$, the eigenstate is of the form
\be \label{eq:FockExpansion}
|\psi^\pm(\ub{P})\rangle=\sum_i z_i^\pm b_{i\pm}^\dagger(\ub{P})|0\rangle
  +\sum_{ijs\lambda}\int d\ub{k} C_{ijs}^{\lambda\pm}(\ub{k})b_{is}^\dagger(\ub{P}-\ub{k})
                                       a_{j\lambda}^\dagger(\ub{k})|0\rangle.
\ee
The normalization condition is
\be  \label{eq:norm}
\langle\psi^{\sigma'}(\ub{P}')|\psi^\sigma(\ub{P})\rangle
                  =\delta(\ub{P}'-\ub{P})\delta_{\sigma'\sigma}.
\ee
For this truncation, we can remove the second PV-fermion flavor,
since it plays no role in the regularization or chiral-symmetry
restoration~\cite{ChiralLimit}.  We let $m_2\rightarrow\infty$,
$\beta_2\rightarrow0$, and $\beta_1\rightarrow1$.

The eigenvalue problem for this state reduces to a set of
coupled equations for the bare-electron amplitudes and
the two-body wave functions:
\bea \label{eq:firstcoupledequation}
[M^2-m_i^2]z_i^\pm &=& \int (P^+)^2 dy d^2k_\perp 
     \sum_{jl\mu}\sqrt{\xi_l}(-1)^{j+l}\epsilon^\lambda e_{l\mu}^{(\lambda)}(\ub{k}) \\
&& \times
  \left[V_{ji\pm}^{\mu*}(\ub{P}-\ub{k},\ub{P})C^{\lambda\pm}_{jl\pm}(\ub{k}) 
        +U_{ji\pm}^{\mu*}(\ub{P}-\ub{k},\ub{P}) C^{\lambda\pm}_{jl\mp}(\ub{k})\right] , \nonumber
\eea
and
\bea  \label{eq:TwoBodyEqns}
\lefteqn{\left[M^2 - \frac{m_j^2 + k_\perp^2}{1-y} - \frac{\mu_{l\lambda}^2 
                                    + k_\perp^2}{y}\right]
  C^{\lambda\pm}_{jls}(\ub{k}) }&&\\
 &&=\sqrt{\xi_l}\sum_{i'} (-1)^{i'} z_{i'}^\pm P^+ e_{l\mu}^{(\lambda)}(\ub{k}) \left[V_{ji'\pm}^\mu(\ub{P}-\ub{k},\ub{P})\delta_{s\pm} 
 +U_{ji'\pm}^\mu(\ub{P}-\ub{k},\ub{P})\delta_{s\mp}\right]
        , \nonumber
\eea
with $y\equiv k^+/P^+$ being the photon's longitudinal momentum fraction.
The second equation (\ref{eq:TwoBodyEqns}) is trivially inverted
to find the two-body wave functions:
\bea \label{eq:TwoBodyWFs}
  C^{\lambda\pm}_{jls}(\ub{k}) &=&
 P^+\sqrt{\xi_l}\sum_{i'} (-1)^{i'} z_{i'}^\pm  e_{l\mu}^{(\lambda)}(\ub{k}) \\
  && \times\frac{\left[V_{ji'\pm}^\mu(\ub{P}-\ub{k},\ub{P})\delta_{s\pm} 
 +U_{ji'\pm}^\mu(\ub{P}-\ub{k},\ub{P})\delta_{s\mp}\right]}
 {\left[M^2 - \frac{m_j^2 + k_\perp^2}{1-y} - \frac{\mu_{l\lambda}^2 
                                    + k_\perp^2}{y}\right]}.
         \nonumber
\eea
Substitution into the first equation (\ref{eq:firstcoupledequation}) yields
a 2$\times$2 matrix eigenvalue problem for the one-body amplitudes $z_i^\pm$:
\bea  \label{eq:2x2}
(M^2-m_i^2)z_i^\pm &=&
      2e^2\sum_{i'} (-1)^{i'}z_{i'}^\pm\left[J+\Delta J+m_im_{i'} (I_0+\Delta I_0)\right. \\
&& \left.-2(m_i+m_{i'}) (I_1+\Delta I_1) \right], \nonumber
\eea
with
\bea \label{eq:In}
I_n(M^2)&=&\int\frac{dy dk_\perp^2}{16\pi^2}
   \sum_{jl}\frac{(-1)^{j+l}\xi_l}{M^2-\frac{m_j^2+k_\perp^2}{1-y}
                                   -\frac{\mu_l^2+k_\perp^2}{y}}
   \frac{m_j^n}{y(1-y)^n}\,, \\
J(M^2)&=&\int\frac{dy dk_\perp^2}{16\pi^2}  \label{eq:J}
   \sum_{jl}\frac{(-1)^{j+l}\xi_l}{M^2-\frac{m_j^2+k_\perp^2}{1-y}
                                   -\frac{\mu_l^2+k_\perp^2}{y}}
   \frac{m_j^2+k_\perp^2}{y(1-y)^2} ,
\eea
and the gauge-dependent parts
\bea \label{eq:DeltaIJ}
\Delta I_0(M^2)&=&-\frac{1-\zeta}{32\pi^2\zeta}\sum_{jl}(-1)^{j+l}\xi_l
\int\frac{dy dk_\perp^2}{y^2(1-y)^2} \\
&& \times 
\frac{m_j^2 y^2+k_\perp^2}
{\left(M^2-\frac{m_j^2+k_\perp^2}{1-y}-\frac{\mu_l^2+k_\perp^2}{y}\right)
\left(M^2-\frac{m_j^2+k_\perp^2}{1-y}-\frac{\tilde\mu_l^2+k_\perp^2}{y}\right)}, 
\nonumber \\
\Delta I_1(M^2)&=&\frac{1-\zeta}{32\pi^2\zeta}\sum_{jl}(-1)^{j+l}\frac{m_j\xi_l}{2}
\int\frac{dy dk_\perp^2}{y^2(1-y)^2} \\
&& \times 
\frac{M^2 y(1-y)-m_j^2 y-k_\perp^2}
{\left(M^2-\frac{m_j^2+k_\perp^2}{1-y}-\frac{\mu_l^2+k_\perp^2}{y}\right)
\left(M^2-\frac{m_j^2+k_\perp^2}{1-y}-\frac{\tilde\mu_l^2+k_\perp^2}{y}\right)}, 
\nonumber \\
\Delta J(M^2)&=&-\frac{1-\zeta}{32\pi^2\zeta}\sum_{jl}(-1)^{j+l}\xi_l
\int\frac{dy dk_\perp^2}{y^2(1-y)^2}  \\
&& \times 
\frac{\left(M^2-\frac{m_j^2}{1-y}\right)^2(1-y)^2+m_j^2k_\perp^2}
{\left(M^2-\frac{m_j^2+k_\perp^2}{1-y}-\frac{\mu_l^2+k_\perp^2}{y}\right)
\left(M^2-\frac{m_j^2+k_\perp^2}{1-y}-\frac{\tilde\mu_l^2+k_\perp^2}{y}\right)}.
\nonumber
\eea
For $\Delta J$ we have taken advantage of the fact that $\sum_{jl}(-1)^{j+l}\xi_l=0$
to simplify the expression by elimination from the numerator terms that
are proportional to the denominator.

For $\mu_0<M-m_0$, there is a line of poles in the $(y,k_\perp^2)$ plane,
in an arc between the points at 
$y=y_\pm\equiv\left[(M^2-m_0^2+\mu_0^2)\pm\sqrt{(M^2-m_0^2+\mu_0^2)^2-4M^2\mu_0^2}\right]/(2M^2)$
on the longitudinal axis.  For $\mu_0=0$, these reduce to $y_-=0$ and $y_+=1-m_0^2/M^2$,
as considered previously~\cite{OnePhotonQED}.  As in that case, we define
integrals with these poles as principal values.  Also, the same considerations
hold for poles associated with denominators containing $\tilde\mu_0$.

\subsection{Analytic solution}

The matrix problem (\ref{eq:2x2}) can be solved analytically, in terms
of the defined integrals.  The solution is facilitated by
the identity $J+\Delta J=M^2 (I_0+\Delta I_0)$, which 
was shown for Feynman gauge in \cite{ChiralLimit} and is
extended to an arbitrary gauge in Appendix~\ref{sec:identity}.

The analytic solutions are
\be \label{eq:OneBosonEigenvalueProb}
\alpha_{\pm}=\frac{(M\pm m_0)(M\pm m_1)}{8\pi (m_1-m_0)
    \left[2 (I_1+\Delta I_1)\pm M (I_0+\Delta I_0)\right]} ,
\ee
with
\be
z_1=\frac{M \pm m_0}{M \pm m_1}z_0 ,
\ee
and $M=m_e$, the physical electron mass.
The solution with the lower sign is the physical one,
because $M=m_0$ when $\alpha_-=0$.  

We fix $\xi_2$ by requiring chiral-symmetry restoration,
that is, $M=0$ for $m_0=0$ when $\alpha_-$ is
equal to the physical value of $\alpha$.  This
implies that $(I_1+\Delta I_1)$ must be zero. 
From earlier work~\cite{ChiralLimit}, we know that
\be
I_1(0)|_{m_0=0}=\frac{m_1}{16\pi^2}\sum_l (-1)^l\xi_l
    \frac{\mu_l^2/m_1^2}{1-\mu_l^2/m_1^2}\ln(\mu_l^2/m_1^2),
\ee
so that we only need to evaluate 
\be
\Delta I_1(0)|_{m_0=0}=-\frac{m_1(1-\zeta)}{64\pi^2\zeta}\sum_l(-1)^l\xi_l
  \int  \frac{(m_1^2 y+k_\perp^2) dy dk_\perp^2 }
      {(m_1^2 y +\mu_l^2 (1-y)+k_\perp^2)(m_1^2 y +\tilde\mu_l^2 (1-y)+k_\perp^2)}.
\ee
Next, we write $k_\perp^2$ in the numerator as $k_\perp^2+m_1^2-m_1^2$
and use the technique in Appendix~\ref{sec:identity} to conclude that the $k_\perp^2+m_i^2$
combination integrates to zero.\footnote{In general, $k_\perp^2+m_i^2$ is replaced
by $M^2(1-y)^2$, but here $M=0$.}  After separation of the denominator into
two terms, we can easily perform the remaining $k_\perp^2$ integration, to
obtain
\be
\Delta I_1(0)|_{m_0=0}=\frac{m_1^3}{64\pi^2}\sum_l\frac{(-1)^l\xi_l}{\mu_l^2}
  \int dy \ln\left(\frac{m_1^2 y+\mu_l^2(1-y)}{m_1^2 y+\tilde\mu_l^2(1-y)}\right).
\ee
The integral over $y$ yields
\be
\Delta I_1(0)|_{m_0=0}=\frac{m_1^3}{64\pi^2}\sum_l(-1)^l\xi_l
\frac{(\zeta-1)m_1^2\ln(m_1^2/\mu_l^2)-(m_1^2-\mu_l^2)\ln\zeta}
    {\zeta(m_1^2-\mu_l^2)(m_1^2-\mu_l^2/\zeta)}.
\ee
Therefore, the constraint from chiral-symmetry restoration is
\be \label{eq:ChiralConstraint}
\frac{m_1}{16\pi^2}\sum_l (-1)^l\xi_l
\left\{\frac{\mu_l^2/m_1^2}{1-\mu_l^2/m_1^2}\ln(\mu_l^2/m_1^2)
 -\frac{m_1^2}{4}\frac{(\zeta-1)m_1^2\ln(m_1^2/\mu_l^2)-(m_1^2-\mu_l^2)\ln\zeta}
   {\zeta(m_1^2-\mu_l^2)(m_1^2-\mu_l^2/\zeta)}\right\}=0.
\ee
Thus, in any covariant gauge, two PV-photon flavors are required
to maintain the chiral limit.

In the limit of infinite mass for the PV electron,
and with use of $\sum_l (-1)^l \xi_l=0$,
the general constraint reduces to
\be
\frac{\zeta-1}{\zeta}\sum_l (-1)^l\xi_l \ln(\mu_l^2/\mu^2)=0,
\ee
with $\mu$ any mass scale.  Given $\xi_0=1$ and $\xi_1=1+\xi_2$,
this is solved by
\be
\xi_2=-\frac{\ln(\mu_0/\mu_1)}{\ln(\mu_2/\mu_1)}>0.
\ee
If the limit $\mu_2\rightarrow\infty$ can be taken, $\xi_2$
is reduced to zero and the second PV photon flavor is
removed.

For the sector-dependent approach, the analogous quantity
to consider in Yukawa theory is ${\cal A}(M^2)$, defined in
Eq.~(B1) of \cite{Karmanov2}.  Combination of various
pieces used in this definition yields, in the notation
used there,
\be
{\cal A}(M^2)=-\frac{1}{64\pi^2}\int \frac{dx dR_\perp^2}{x(1-x)}
\sum_{ij} 
\frac{(-1)^{i+j} m_i}{\frac{m_i^2+R_\perp^2}{1-x}
                           +\frac{\mu_j^2+R_\perp^2}{x}-M^2}.
\ee
Comparison with our (\ref{eq:In}) shows that 
${\cal A}(M^2)=\frac14 I_1(M^2)|_{\xi_2=0}$.
When $M=0$ and $m_0=0$, the $i=1$ term of ${\cal A}$ is not
zero, unless $m_1\rightarrow\infty$.  Therefore, $\delta m_2$ in
(29b) of \cite{Karmanov2} is also not zero,
in contradiction of chiral-symmetry restoration.
Thus, the sector-dependent approach requires a second PV-photon
flavor in Yukawa theory.  For QED, the situation is not
materially different.

To complete the analysis of the eigensolution, we need to
consider the gauge dependence and infrared dependence of 
the integral combination
\be
\Delta\equiv 2\Delta I_1(M^2)-M \Delta I_0(M^2),
\ee
which enters the denominator of (\ref{eq:OneBosonEigenvalueProb}).
The combination $2I_1(M^2)-M I_0(M^2)$ also appears there,
but is gauge-independent by definition and is
known to be infrared safe~\cite{OnePhotonQED}.  From
the definitions (\ref{eq:DeltaIJ}) of 
$\Delta I_0$ and $\Delta I_1$, we can obtain,
after eliminating $k_\perp^2$ from the numerator
in the same manner as before,
\be
\Delta=\frac{1-\zeta}{32\pi^2\zeta}\sum_{jl}(-1)^{j+l}\xi_l
    \int \frac{dy dk_\perp^2}{y^2(1-y)}
    \frac{(M-m_j)^2[M(1-y)+m_j]}
    {\left(M^2-\frac{m_j^2+k_\perp^2}{1-y}-\frac{\mu_l^2+k_\perp^2}{y}\right)
      \left(M^2-\frac{m_j^2+k_\perp^2}{1-y}-\frac{\tilde\mu_l^2+k_\perp^2}{y}\right)}.
\ee
The $k_\perp^2$ integral yields
\bea \label{eq:yDelta}
\Delta&=&\frac{1}{32\pi^2}\sum_{jl}(-1)^{j+l}\frac{\xi_l}{\mu_l^2}(M-m_j)^2 \\
  && \times\int dy [M(1-y)+m_j] 
      \ln\left\{\left|\frac{m_j^2 y+\mu_l^2(1-y)-M^2 y(1-y)}
                         {m_j^2 y+\tilde\mu_l^2(1-y)-M^2 y(1-y)}\right|\right\}.
      \nonumber
\eea
The $j=0$ term is of order $(M-m_0)^2\propto\alpha^2$, and the
$j=1$ term is of order $1/m_1$, after invocation of the chiral
constraint (\ref{eq:ChiralConstraint}) to eliminate the leading,
order-$m_1$ term.  Thus, $\Delta$ breaks gauge invariance only
in ways to be expected; gauge invariance can be attained only without
truncations and then only in the $m_1\rightarrow\infty$ limit.  The
high-order $\alpha$ correction is a signal of a truncation effect,
and, of course, the $1/m_1$ contribution disappears as $m_1\rightarrow\infty$.

To study the dependence on the IR mass scale $\mu_0$, we consider
the $\mu_0\rightarrow0$ limit of the $j=l=0$ term in (\ref{eq:yDelta}),
which is
\be
\frac{\zeta-1}{32\pi^2\zeta}(M-m_0)^2\int_0^1 dy \frac{M(1-y)+m_0}{y[m_0^2-M^2(1-y)]}.
\ee
The behavior near $y=0$ is such that the term has a log divergence
multiplied by $(\zeta-1)(M-m_0)$.  Thus, this contribution is of 
order $\alpha$, not $\alpha^2$, and, of course, is absent in Feynman
gauge.  The order of the contribution is, however, still consistent
with being a truncation error.  Also, the presence of an IR divergence
is consistent with the presence of other, UV divergences, which
are uncanceled due to truncation, as discussed in \cite{OnePhotonQED}
and as discussed below, with respect to normalization of the eigenstate.

\subsection{Anomalous magnetic moment}

We compute the anomalous magnetic moment of the dressed electron
from the spin-flip matrix element of the electromagnetic current
$J$~\cite{BrodskyDrell}.  The plus component of the current is
used because, in the absence of vacuum polarization, it is not
renormalized~\cite{BRS,ChiralLimit}.  In general, the transition
amplitude for absorption of a photon of momentum $q$ by a dressed
electron is given by
\be
\langle\psi^{\sigma}(\ub{P}+\ub{q})|\frac{J^+(0)}{P^+}|\psi^\pm(\ub{P})\rangle
=2\delta_{\sigma\pm}F_1(q^2)\pm\frac{q^1\pm iq^2}{M}\delta_{\sigma\mp}F_2(q^2),
\ee
where $F_1$ and $F_2$ are the usual Dirac and Pauli form factors.
The anomalous moment is $a_e=F_2(0)$; normalization of the state
is equivalent to $F_1(0)=1$.  As described in \cite{BrodskyDrell},
the limit of zero momentum transfer for $F_2$ can be written as
\be \label{eq:ae}
a_e=\mp M\sum_{jls\lambda}\int d\ub{k} \epsilon^\lambda (-1)^{j+l}
  C_{jls}^{\lambda\pm *}(\ub{k})
  y\left(\frac{\partial}{\partial k^1}\pm i\frac{\partial}{\partial k^2}\right)
  C_{jls}^{\lambda\mp}(\ub{k}).
\ee
This form assumes complete separation of the internal and external
momentum variables in the wave functions $C_{jls}^{\lambda\pm}$,
which does occur for components in terms of polarizations.  
The sum over polarizations $\lambda$ does not include the gauge
projection, because gauge invariance has already been broken
by both the truncation and the flavor-changing currents.
The normalization condition (\ref{eq:norm}), or $F_1(0)=1$, becomes
\be  \label{eq:norm2}
1=(z_0^\pm)^2-(z_1^\pm)^2+\sum_{jls\lambda}\int d\ub{k} 
   \epsilon^\lambda (-1)^{j+l} |C_{jls}^{\lambda\pm}(\ub{k})|^2,
\ee
which determines $z_0^\pm$.

\begin{figure}[ht]
\vspace{0.2in}
\centerline{\includegraphics[width=11cm]{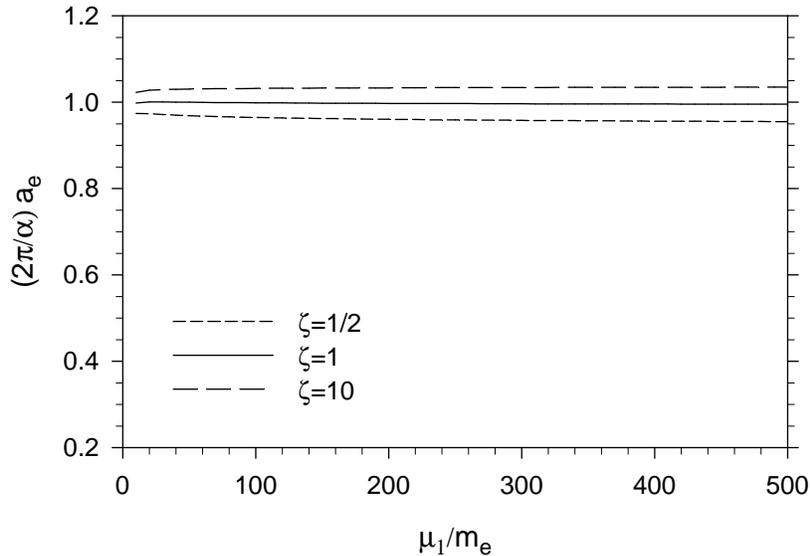}}
\caption{\label{fig:aevsmu1} 
The anomalous magnetic moment $a_e$ for the dressed-electron
state truncated to include at most the one-electron/one-photon
Fock states, as a function of the PV-photon mass $\mu_1$.
The IR mass scale $\mu_0$ is $0.001\,m_e$, and the gauge
parameter $\zeta$ is 1/2, 1, and 10.
}
\end{figure}

\begin{figure}[ht]
\vspace{0.2in}
\centerline{\includegraphics[width=11cm]{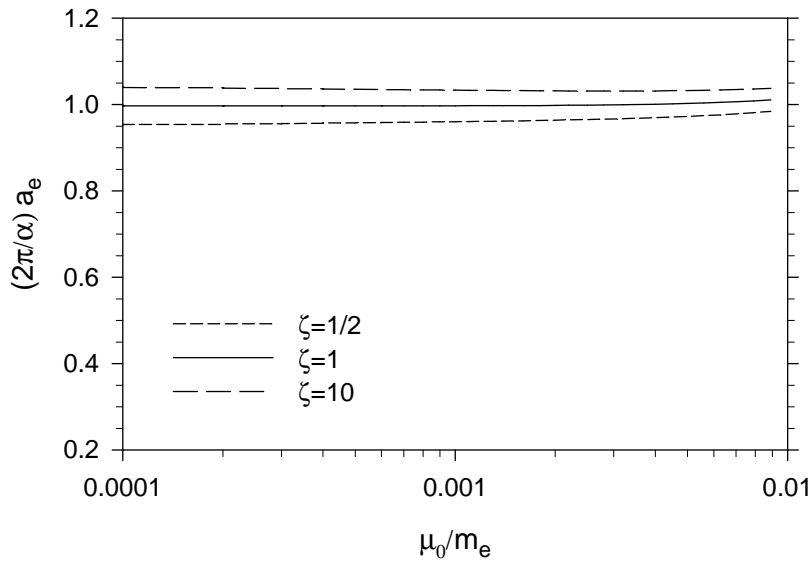}}
\caption{\label{fig:aevsmu0} 
Same as Fig.~\ref{fig:aevsmu1} but as a function of the 
IR mass scale $\mu_0$, with $\mu_1=200\,m_e$.
}
\end{figure}

\begin{figure}[ht]
\vspace{0.2in}
\centerline{\includegraphics[width=11cm]{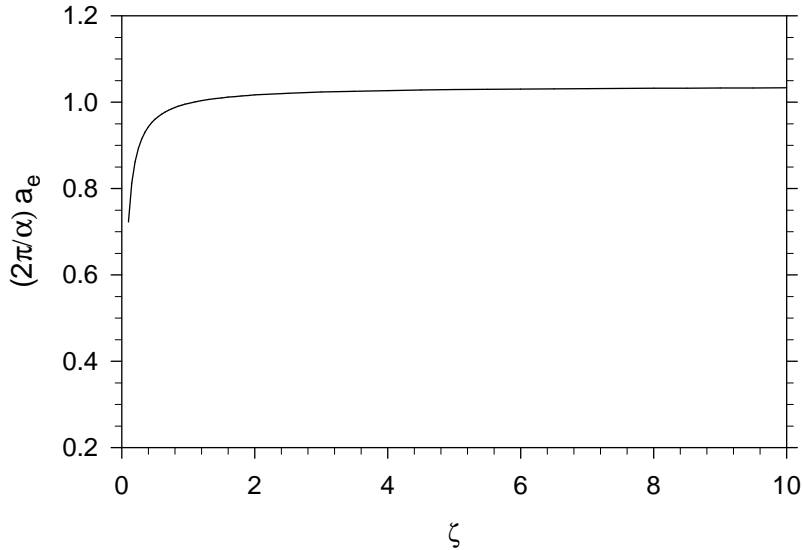}}
\caption{\label{fig:aevszeta} 
Same as Fig.~\ref{fig:aevsmu1} but as a function of the
gauge parameter $\zeta$, with $\mu_0=0.001\,m_e$ and $\mu_1=200\,m_e$.
}
\end{figure}

The reduction of the expression for the anomalous moment is
given in Appendix~\ref{sec:anomalousmoment}.  The result in
the limit of infinite PV electron mass is given in (\ref{eq:ae-m1infty}).
The normalization condition is evaluated in Appendix~\ref{sec:normalization},
with the same infinite mass limit yielding (\ref{eq:m1limit}).
From these expressions, the anomalous moment can be computed
for various values of the UV scale $\mu_1$, the IR scale $\mu_0$,
and the gauge parameter $\zeta$.  Sample results are given in
Figs.~\ref{fig:aevsmu1}, \ref{fig:aevsmu0}, and \ref{fig:aevszeta}.
%%%%%%%%%%%%%%%%%%%%%%%%%%%%%%%%%%%%%
%%%%%%%%%%%%%%%%%%%%%%%%%%%%%%%%%%%%%%%%%%%%%%%%%%%%%%%%%%%%%%
%%%%%%%%%%%%%%%%%%%%%%%%%%%%%%%%%%%%%
%%%%%%%%%%%%%%%%%%%%%%%%%%%%%%%%%%%%%%%%%%%%%%%%%%%%%%%%%%%%%%
%%%%%%%%%%%%%%%%%%%%%%%%%%%%%%%%%%%%%
%%%%%%%%%%%%%%%%%%%%%%%%%%%%%%%%%%%%%%%%%%%%%%%%%%%%%%%%%%%%%%
Because truncation errors do not allow either $\mu_0\rightarrow0$,
except in Feynman gauge, or $\mu_1\rightarrow\infty$,
we look for regions in each where the physical quantity
is relatively flat.  In Fig.~\ref{fig:aevszeta}, we then
look for sensitivity to the gauge parameter when
$\mu_0$ and $\mu_1$ have values in such regions.  The
plot shows little sensitivity and,
therefore, approximate gauge independence, except for small values
of $\zeta$.  For small $\zeta$, the theory is near the singular
limit where the gauge-fixing term is removed from the Lagrangian
(\ref{eq:Lagrangian}) and truncation errors are amplified.

The remaining gauge dependence can be seen to be consistent
with the order of the truncation in the calculation.
Because $M-m_0$ is of order $\alpha$, we have,
for the leading term in $a_e$, as given by (\ref{eq:ae-m1infty}),
\be
M\int_0^1 dy \frac{y(1-y)My}{m_0^2y-M^2 y(1-y)}=\frac12+{\cal O}(\alpha),
\ee
and, therefore,
\be
a_e=\frac{\alpha}{2\pi}z_0^2+{\cal O}(\alpha^2,1/\mu_1^2).
\ee
Up to normalization, the Schwinger result~\cite{Schwinger} of $\alpha/2\pi$
is recovered, and the gauge-dependent contributions, as well as some 
physical nonperturbative contributions, are higher order in $\alpha$, 
consistent with the truncation to one photon in the Fock basis.

To complete the analysis, we must consider the 
normalization factor $z_0$, as given in (\ref{eq:m1limit}).
At the present order of truncation, a value other
than 1 for $z_0$ represents a truncation error, in the sense
that contributions occurring at the same order in $\alpha$
for the numerator of the expectation value for the
anomalous moment have been left out by the truncation; therefore,
the gauge dependence of $z_0$ must be due to truncation errors.

In the IR limit $\mu_0\rightarrow0$, the normalization $z_0$ does have
a singular contribution of the form 
$3\frac{\alpha}{4\pi}\frac{1-\zeta}{\zeta}\int \frac{dy}{y}$.  This
comes from a combination of the last two terms of the curly bracket in
(\ref{eq:m1limit}) for $\mu_0\rightarrow0$ and $l=0$.  Therefore,
the normalization contains an IR divergence in addition to its usual
UV divergence, except in Feynman gauge, where there is only a
UV divergence.

These divergences are the characteristic ``uncanceled divergences''
caused by Fock-space truncation~\cite{OnePhotonQED}.  They
arise in both the standard and sector-dependent parameterizations,
although in the latter case the IR divergence is present
even in Feynman gauge~\cite{SecDep}.  For the standard
parameterization used here, we find the divergences in the
normalization factor.  For the sector-dependent parameterization,
the divergence is most easily seen by considering the probability
for the one-electron/one-photon sector.  This should be between zero and one,
but the renormalization of the sector-dependent coupling absorbs
the divergence in the normalization factor $1/z_0^2$ and allows
the probability of the one-electron/one-photon sector to diverge.
In this case, the overall norm of the eigenstate is maintained only because
the probability for the bare-electron sector goes to negative
infinity in such a way that the sum of probabilities is formally one.

Because of the uncanceled divergences, not all of the PV masses can
be taken to infinity and, except for our standard parameterization 
in Feynman gauge, the physical photon mass cannot be taken to zero.
As argued in \cite{OnePhotonQED}, the errors introduced by these
limitations are to be minimized by seeking ranges of mass values
over which results do not change significantly.  This strikes a
balance between the errors caused by the presence of unphysical
PV fields and a nonzero photon mass and the errors associated
with Fock-space truncation.  The former decrease with increasing
PV masses (as the PV fields are removed from the spectrum) and 
decreasing photon mass; the latter, the truncation errors, increase
with increasing PV masses as the uncanceled divergences assert
themselves.

%%%%%%%%%%%%%%%%%%%%%%
\section{Summary}
\label{sec:Summary}
%%%%%%%%%%%%%%%%%%%%%%

We have developed a formalism whereby nonperturbative calculations
can be done for light-front QED in an arbitrary covariant gauge.
The formalism combines a light-front Stueckelberg quantization
for the free photon field with a Pauli--Villars regularization
that simplifies the constraint equation for the nondynamical
part of the fermion field.  The Stueckelberg quantization allows the
physical and PV photons to be handled in the same way,
which facilitates the regularization and the preservation of
symmetries.  In Feynman gauge, this quantization is equivalent
to the Gupta--Bleuler quantization used
previously~\cite{OnePhotonQED,ChiralLimit,TwoPhotonQED}.

As a first application of the formalism, we have
studied the dressed electron in a Fock space truncated
to include at most one photon and no positrons.  In
particular, we have investigated the gauge invariance
of the mass shift and the anomalous magnetic moment.
In both cases, the residual gauge dependence can be
ascribed to errors induced by the Fock-space truncation.
The dependence of the anomalous moment on the gauge parameter
$\zeta$ is illustrated in Fig.~\ref{fig:aevszeta} in the
case where the PV fermion mass $m_1$ is infinite.
If the PV-fermion mass is kept finite, there is
also gauge dependence of order $1/m_1$, due to
fermion-flavor-mixing currents.  The strong dependence
as $\zeta\rightarrow0$ is to be expected, because in
this limit the gauge-fixing term is removed from the
Lagrangian and the theory becomes undefined.  We have also
found that two PV-photon flavors are required to maintain
the chiral symmetry of the massless-electron limit;
this extends the earlier Feynman-gauge result~\cite{ChiralLimit}
to arbitrary gauges.

The formalism can be applied to higher-order truncations
and to other charge sectors.  For high-order truncations,
where calculations are done numerically~\cite{TwoPhotonQED},
the PV-fermion mass $m_1$ must usually be kept finite; however,
with $m_1$ large enough, the gauge-dependent effects
should be small.

\acknowledgments
This work was supported in part by the Department of Energy
through Contract No.\ DE-FG02-98ER41087
and by the Minnesota Supercomputing Institute through
grants of computing time.

\appendix

%%%%%%%%%%%%%%%%%%%%%%%%%%%%%%%%%%%%%%%%%
\section{An integral identity}  \label{sec:identity}
%%%%%%%%%%%%%%%%%%%%%%%%%%%%%%%%%%%%%%%%%%%

From \cite{ChiralLimit}, we have already that $J=M^2 I_0$.
Thus, to show that $J+\Delta J=M^2 (I_0+\Delta I_0)$,
we need only consider the gauge-dependent parts.  If
we write the common denominators in (\ref{eq:DeltaIJ})
as the difference of two terms, we obtain
\bea \label{eq:DeltaJ-DeltaI0}
\Delta J-M^2\Delta I_0&=&
-\frac{1}{32\pi^2}\sum_{jl}(-1)^{j+l}\xi_l
  \frac{M^2-m_j^2}{\mu_l^2}\int\frac{dy dk_\perp^2}{y}
  \left[M^2-\frac{m_j^2+k_\perp^2}{(1-y)^2}\right] \\
 &&\times \left(\frac{1}{M^2-\frac{m_j^2+k_\perp^2}{1-y}-\frac{\mu_l^2+k_\perp^2}{y}}
  -\frac{1}{M^2-\frac{m_j^2+k_\perp^2}{1-y}-\frac{\tilde\mu_l^2+k_\perp^2}{y}}\right),
  \nonumber
\eea
where we have used $1/(\tilde\mu_l^2-\mu_l^2)=\zeta/(1-\zeta)\mu_l^2$ to
simplify the leading factors.  For the terms that contain $m_j^2+k_\perp^2$,
we change the integration variable $y$ to
\be \label{eq:x-variable}
x=(1-y)\frac{\mu_l^2+k_\perp^2}{m_j^2 y +\mu_l^2(1-y)+k_\perp^2}
\ee
or
\be
\tilde x=(1-y)\frac{\tilde\mu_l^2+k_\perp^2}{m_j^2 y +\tilde\mu_l^2(1-y)+k_\perp^2},
\ee
depending on whether $\mu_l$ or $\tilde\mu_l$ appears in the
denominator of the integrand.  As discussed in \cite{ChiralLimit},
these variables range between 0 and 1, though in reverse order,
and satisfy
\be
\frac{m_j^2+k_\perp^2}{1-y}+\frac{\mu_l^2+k_\perp^2}{y}
  =\frac{m_j^2+k_\perp^2}{1-x}+\frac{\mu_l^2+k_\perp^2}{x},
\ee
and the analogous expression for $\tilde x$.  With use of this
identity and differentiation of (\ref{eq:x-variable}),
we obtain
\be
(m_j^2+k_\perp^2)\frac{dy}{y(1-y)^2}
   =\frac{dx}{x}\left(M^2-\frac{m_j^2+k_\perp^2}{1-x}
       -\frac{\mu_l^2+k_\perp^2}{x}-M^2\right).
\ee
The analogous expression holds for $d\tilde x$.  Substitution
into (\ref{eq:DeltaJ-DeltaI0}) and replacement of $x$ and $\tilde x$ by $y$,
to have a common integration variable for all terms, yields
\bea
\Delta J-M^2\Delta I_0&=&
-\frac{1}{32\pi^2}\sum_{jl}(-1)^{j+l}\xi_l
  \frac{M^2-m_j^2}{\mu_l^2}\int\frac{dy dk_\perp^2}{y} \\
 && \times
 \left\{M^2\left(\frac{1}{M^2-\frac{m_j^2+k_\perp^2}{1-y}-\frac{\mu_l^2+k_\perp^2}{y}}
  -\frac{1}{M^2-\frac{m_j^2+k_\perp^2}{1-y}-\frac{\tilde\mu_l^2+k_\perp^2}{y}}\right)\right. 
  \nonumber \\
 && \left.
  +\left[\left(1-\frac{M^2}{M^2-\frac{m_j^2+k_\perp^2}{1-y}-\frac{\mu_l^2+k_\perp^2}{y}}\right)
  -\left(1-\frac{M^2}{M^2-\frac{m_j^2+k_\perp^2}{1-y}-\frac{\tilde\mu_l^2+k_\perp^2}{y}}\right)
       \right]  \right\} .  \nonumber
\eea
Here we have taken into account the reversed order of limits for $x$ and $\tilde x$ by changing
the sign of the terms in the square brackets.  Clearly, the sum of terms in the
curly brackets is zero, and, therefore, $\Delta J=M^2 \Delta I_0$.

%%%%%%%%%%%%%%%%%%%%%%%%%%%%%%%%%%%%%%%%%%%%%%%%%%%%%%
\section{Evaluation of the anomalous-moment formula} \label{sec:anomalousmoment}
%%%%%%%%%%%%%%%%%%%%%%%%%%%%%%%%%%%%%%%%%%%%%%%%%%%%

On substitution of the two-body wave functions (\ref{eq:TwoBodyWFs})
and use of the Kronecker deltas in spin, the expression (\ref{eq:ae})
for the anomalous moment becomes
\bea \label{eq:aedef}
a_e&=&- M\sum_{jl\lambda}\sum_{i'i''} \epsilon^\lambda (-1)^{j+l+i'+i''}
    \xi_l z_{i'}^+ z_{i''}^- \int dy d k_\perp^2 
    \frac{y \pi (P^+)^3}
    {M^2 - \frac{m_j^2 + k_\perp^2}{1-y} - \frac{\mu_{l\lambda}^2+k_\perp^2}{y}}   \\
&& \times    \left\{
    e_{l\mu'}^{(\lambda)}(\ub{k}) V_{ji'+}^{\mu' *}(\ub{P}-\ub{k},\ub{P})
    \left(\frac{\partial}{\partial k^1}+ i\frac{\partial}{\partial k^2}\right)
    \frac{e_{l\mu}^{(\lambda)}(\ub{k}) U_{ji''-}^\mu(\ub{P}-\ub{k},\ub{P})}
    {M^2 - \frac{m_j^2 + k_\perp^2}{1-y} - \frac{\mu_{l\lambda}^2+k_\perp^2}{y}} 
          \right. \nonumber \\
&&  \left.
    +e_{l\mu'}^{(\lambda)}(\ub{k})U_{ji'+}^{\mu' *}(\ub{P}-\ub{k},\ub{P})
    \left(\frac{\partial}{\partial k^1}+ i\frac{\partial}{\partial k^2}\right)
    \frac{e_{l\mu}^{(\lambda)}(\ub{k}) V_{ji''-}^\mu(\ub{P}-\ub{k},\ub{P})}
    {M^2 - \frac{m_j^2 + k_\perp^2}{1-y} - \frac{\mu_{l\lambda}^2+k_\perp^2}{y}}
    \right\}, \nonumber
\eea
with the vertex functions specified in (\ref{eq:vertices}) and (\ref{eq:vertices2}).
The terms generated by differentiation of the denominators cancel.  Simplification
of the remaining terms, summed over polarizations $\lambda$, yields
\bea
a_e&=&-M \frac{e^2}{8\pi^2}\sum_{jl}\sum_{i'i''}(-1)^{j+l+i'+i''}\xi_l z_{i'}^+ z_{i''}^- 
     \int dy dk_\perp^2      \\
&& \times \left\{ 
\frac{m_{i'}(1-y)-m_j}{y(1-y)}
\frac{2}{\left(M^2 - \frac{m_j^2 + k_\perp^2}{1-y} - \frac{\mu_l^2+k_\perp^2}{y}\right)^2}
+\frac{m_{i''}-m_j}{y(1-y)}
  \frac{2+(1-\zeta)/\zeta}
  {\left(M^2 - \frac{m_j^2 + k_\perp^2}{1-y} - \frac{\mu_l^2+k_\perp^2}{y}\right)^2} 
     \right. \nonumber \\
 && \rule{0.5in}{0mm}
 + \frac{m_{i''}-m_j}{\mu_l^2(1-y)}\left(\frac{m_j m_{i'}y}{1-y}+\frac{\mu_l^2}{y\zeta}\right)
  \nonumber \\
 && \rule{1in}{0mm} \left. \times
 \left(\frac{1}{\left(M^2 - \frac{m_j^2 + k_\perp^2}{1-y} 
                       - \frac{\tilde\mu_l^2+k_\perp^2}{y}\right)^2}
-\frac{1}{\left(M^2 - \frac{m_j^2 + k_\perp^2}{1-y} - \frac{\mu_l^2+k_\perp^2}{y}\right)^2}
   \right)\right\}. \nonumber
   \eea
The $k_\perp^2$ integrals have double poles, if $\mu_0<M-m_0$ or $\tilde\mu_0<M-m_0$;
following the earlier convention in \cite{OnePhotonQED}, we define the integrals by
\be \label{eq:dblepole}
\int \frac{f(x)dx}{(x-a)^2}=\lim_{\eta\rightarrow0}\frac{1}{2\eta}
\left[{\cal P}\int\frac{f(x)dx}{x-a-\eta}-{\cal P}\int\frac{f(x)dx}{x-a+\eta}\right].
\ee
With or without the pole, we find
\be  \label{eq:dk2}
\int\frac{dk_\perp^2}
{\left(M^2 - \frac{m_j^2 + k_\perp^2}{1-y} - \frac{\mu_l^2+k_\perp^2}{y}\right)^2} 
=\frac{y^2 (1-y)^2}{m_j^2 y +\mu_l^2 (1-y)-M^2 y(1-y)}.
\ee
These leave the anomalous moment in the form
\be
a_e=\frac{\alpha}{\pi}M\sum_{jl}\sum_{i'i''}(-1)^{j+l+i'+i''}\xi_l z_{i'}^+z_{i''}^-
\int dy \frac{y(1-y)[m_{i'}y-(m_{i'}+m_{i''}-2m_j)]}
       {m_j^2 y + \mu_l^2 (1-y)-M^2 y(1-y)} + \Delta a_e,
\ee
with $\Delta a_e$ a gauge-dependent part given by
\bea
\lefteqn{\Delta a_e=-\frac{1-\zeta}{\zeta}\frac{\alpha}{2\pi}M\sum_{jl} \sum_{i'i''}
(-1)^{j+l+i'+i''}\xi_l z_{i'}^+z_{i''}^- (m_{i''}-m_j)\int dy y(1-y)}&& \nonumber \\
&& \times \left\{\frac{1}{m_j^2 y+\mu_l^2 (1-y)-M^2 y(1-y)} \right.  \\
&& \left. -\frac{m_j m_{i'} y^2 +\mu_l^2(1-y)/\zeta}
      {[m_j^2 y + \mu_l^2 (1-y)-M^2 y(1-y)]
                  [m_j^2 y + \tilde\mu_l^2 (1-y)-M^2 y(1-y)]}\right\}. \nonumber
\eea

Except for the normalization factors $z_i^\pm$, this expression
is IR safe; the photon mass $\mu_0$ can be set to zero.  We first take
the limit $m_1\rightarrow\infty$, which, with use of $z_i^+=z_i^-\equiv z_i$,
implies $\sum_i(-1)^iz_i\rightarrow z_0$ and $\sum_i (-1)^i m_i z_i\rightarrow Mz_0$.
Next, the second PV-photon flavor is removed by the limit $\mu_2\rightarrow\infty$,
in which $\xi_2\rightarrow 0$ and $\xi_1\rightarrow 1$, to obtain
\be \label{eq:ae-m1infty}
a_e\rightarrow\frac{\alpha}{\pi}Mz_0^2\sum_{l=0}^1 (-1)^l \int dy
\frac{y(1-y)[My-2(M-m_0)]}{m_0^2 y +\mu_l^2 (1-y)-M^2 y(1-y)}+\Delta a_e
\ee
and
\bea
\lefteqn{\Delta a_e\rightarrow -\frac{1-\zeta}{\zeta}\frac{\alpha}{2\pi}M(M-m_0)z_0^2
\sum_{l=0}^1 (-1)^l\int dy
\left\{\frac{y(1-y)}{m_0^2 y +\mu_l^2 (1-y)-M^2 y(1-y)} \right. } && \nonumber \\
&&\left. -\frac{y(1-y)[m_0My^2+\mu_l^2(1-y)/\zeta]}
    {[m_0^2 y + \mu_l^2 (1-y)-M^2 y(1-y)]
                  [m_0^2 y + \tilde\mu_l^2 (1-y)-M^2 y(1-y)]}\right\}. 
\eea
%

%%%%%%%%%%%%%%%%%%%%%%%%%%%%%%%%%%%%%%%%%%%%%%%%%%%%%%
\section{Evaluation of the normalization condition} \label{sec:normalization}
%%%%%%%%%%%%%%%%%%%%%%%%%%%%%%%%%%%%%%%%%%%%%%%%%%%%

On substitution of the two-body wave functions (\ref{eq:TwoBodyWFs}),
the normalization condition (\ref{eq:norm2}) becomes
\bea
1&=&(z_0^\pm)^2-(z_1^\pm)^2
+\sum_{jl}\sum_{i'i''}(-1)^{j+l+i'+i''}\xi_l z_{i'}^\pm z_{i''}^\pm 
\sum_\lambda \epsilon^\lambda \int 
\frac{\pi (P^+)^3 dy dk_\perp^2}
{\left( M^2-\frac{m_j^2+k_\perp^2}{1-y}-\frac{\mu_{l\lambda}^2+k_\perp^2}{y}\right)^2} \\
&& \times e_{l\mu'}^{(\lambda)}(\ub{k}) e_{l\mu}^{(\lambda)}(\ub{k})
\left[ V_{ji'\pm}^{\mu' *}(\ub{P}-\ub{k},\ub{P})V_{ji''\pm}^\mu(\ub{P}-\ub{k},\ub{P})
      +U_{ji'\pm}^{\mu' *}(\ub{P}-\ub{k},\ub{P})U_{ji''\pm}^\mu(\ub{P}-\ub{k},\ub{P})
         \right].   \nonumber
\eea
To simplify the expression, we first add and subtract the $\lambda=0$
term with the denominator replaced by the denominator of the $\lambda\neq0$
terms.  We then have
\bea
1&=&(z_0^\pm)^2-(z_1^\pm)^2
+\sum_{jl}\sum_{i'i''}(-1)^{j+l+i'+i''}\xi_l z_{i'}^\pm z_{i''}^\pm 
\int \pi (P^+)^3 dy dk_\perp^2 \\
&& \times \left\{
   \frac{1}
   {\left( M^2-\frac{m_j^2+k_\perp^2}{1-y}-\frac{\mu_l^2+k_\perp^2}{y}\right)^2}
   \sum_\lambda \epsilon^\lambda e_{l\mu'}^{(\lambda)}(\ub{k}) e_{l\mu}^{(\lambda)}(\ub{k}) 
   \right. \nonumber \\
&& \left. +  \left(
   \frac{1}
   {\left( M^2-\frac{m_j^2+k_\perp^2}{1-y}-\frac{\mu_l^2+k_\perp^2}{y}\right)^2}
-   \frac{1}
   {\left( M^2-\frac{m_j^2+k_\perp^2}{1-y}-\frac{\tilde\mu_l^2+k_\perp^2}{y}\right)^2}
   \right)e_{l\mu'}^{(0)}(\ub{k}) e_{l\mu}^{(0)}(\ub{k})\right\} \nonumber \\
&& \times \left[ V_{ji'\pm}^{\mu' *}(\ub{P}-\ub{k},\ub{P})V_{ji''\pm}^\mu(\ub{P}-\ub{k},\ub{P})
      +U_{ji'\pm}^{\mu' *}(\ub{P}-\ub{k},\ub{P})U_{ji''\pm}^\mu(\ub{P}-\ub{k},\ub{P})
         \right].   \nonumber
\eea
On use of the vertex functions in (\ref{eq:vertices}) and (\ref{eq:vertices2})
and of $z_i^+=z_i^-\equiv z_i$, we obtain,
\bea
1&=&z_0^2-z_1^2+\frac{e^2}{16\pi^2}\sum_{jl}\sum_{i'i''}(-1)^{j+l+i'+i''}\xi_l z_{i'}z_{i''}
  \int dy dk_\perp^2 \\
&& \left\{\frac{1}
    {\left( M^2-\frac{m_j^2+k_\perp^2}{1-y}-\frac{\mu_l^2+k_\perp^2}{y}\right)^2}
    \left[ 2\frac{m_j^2-2m_j(m_{i'}+m_{i''})(1-y)+m_{i'}m_{i''}(1-y)^2+k_\perp^2}{y(1-y)^2}
    \right. \right. \nonumber \\
 && \rule{2in}{0mm} \left.  -\frac{1-\zeta}{\zeta}\left(\frac{m_j(m_{i'}+m_{i''})}{y(1-y)}
                                      +\frac{2k_\perp^2}{y^3(1-y)}
                                      +\frac{1+\zeta}{\zeta}\frac{\mu_l^2}{y^3}\right)\right] 
             \nonumber \\
&&  +\frac{1}{\mu_l^2}\left[
    \frac{1}
    {\left( M^2-\frac{m_j^2+k_\perp^2}{1-y}-\frac{\mu_l^2+k_\perp^2}{y}\right)^2}
    -\frac{1}
    {\left( M^2-\frac{m_j^2+k_\perp^2}{1-y}-\frac{\tilde\mu_l^2+k_\perp^2}{y}\right)^2}
    \right]  \nonumber \\
&&  \times
    \left[ \frac{m_j^2 m_{i'} m_{i''} y}{(1-y)^2}
           +\frac{k_\perp^4}{y^3(1-y)^2}
           +\frac{\tilde\mu_l^4}{y^3}
           +\frac{(m_j^2+m_{i'}m_{i''})k_\perp^2}{y(1-y)^2}
           \right. \nonumber \\
&& \rule{2in}{0mm}  \left. \left.
           +\frac{\tilde\mu_l^2 m_j(m_{i'}+m_{i''})}{y(1-y)}
           +\frac{2\tilde\mu_l^2 k_\perp^2}{y^3(1-y)}\right]
             \rule{0mm}{0.4in} \right\}. \nonumber
\eea
The $k_\perp^2$ integrals are defined by (\ref{eq:dblepole}) when a
pole is present.  The integrals needed are given by (\ref{eq:dk2}),
\be
\int\frac{k_\perp^2 dk_\perp^2}
{\left( M^2-\frac{m_j^2+k_\perp^2}{1-y}-\frac{\mu_l^2+k_\perp^2}{y}\right)^2}
=-y^2(1-y)^2 \ln[|m_j^2 y+\mu_l^2(1-y)-M^2 y(1-y)|],
\ee
and
\bea
\int\frac{k_\perp^4 dk_\perp^2}
{\left( M^2-\frac{m_j^2+k_\perp^2}{1-y}-\frac{\mu_l^2+k_\perp^2}{y}\right)^2}
&=&y^2(1-y)^2\left(m_j^2 y+\mu_l^2(1-y)-M^2 y(1-y)\right) \\
&& \times \left(2\ln[|m_j^2 y+\mu_l^2(1-y)-M^2 y(1-y)|]+1\right),  \nonumber 
\eea
where we have dropped infinite terms that cancel in the final
expressions, due to either $\sum_l(-1)^l\xi_l=0$ or the difference
between the two denominators, 
$(M^2-\frac{m_j^2+k_\perp^2}{1-y}-\frac{\mu_l^2+k_\perp^2}{y})^2$
and $(M^2-\frac{m_j^2+k_\perp^2}{1-y}-\frac{\tilde\mu_l^2+k_\perp^2}{y})^2$.
Substitution of these integrals yields
\bea  \label{eq:rawnorm}                  
1&=&z_0^2-z_1^2+\frac{e^2}{16\pi^2}\sum_{jl}\sum_{i'i''}(-1)^{j+l+i'+i''}\xi_l z_{i'}z_{i''}
  \int dy \\
&& \left\{ \frac{1}{m_j^2 y+\mu_l^2(1-y)-M^2 y(1-y)}
\left[\rule{0mm}{0.3in} 2y(m_j^2-2m_j(m_{i'}+m_{i''})(1-y)+m_{i'}m_{i''}(1-y)^2) 
                      \right.  \right. \nonumber \\
&& \rule{1in}{0mm}  \left.
      -\frac{1-\zeta}{\zeta}\left(m_j(m_{i'}+m_{i''})y(1-y)
                    +\frac{1+\zeta}{\zeta}\frac{\mu_l^2(1-y)^2}{y}
                    \right)\right] \nonumber \\
&& -2\left(y-\frac{1-\zeta}{\zeta}\frac{1-y}{y}\right)
        \ln[|m_j^2 y+\mu_l^2(1-y)-M^2 y(1-y)|] \nonumber \\
&& +\frac{1}{\mu_l^2}\left[\frac{1}{m_j^2 y+\mu_l^2(1-y)-M^2 y(1-y)}
                          -\frac{1}{m_j^2 y+\tilde\mu_l^2(1-y)-M^2 y(1-y)}\right] \nonumber \\
&& \rule{1in}{0mm} \times 
             \left[ m_j^2 m_{i'}m_{i''}y^3+\frac{\tilde\mu_l^4(1-y)^2}{y}
                         +\tilde\mu_l^2 m_i (m_{i'}+m_{i''})y(1-y)\right] \nonumber \\
&& +\frac{1}{\mu_l^2 y}
\left[(m_j^2 y+\mu_l^2(1-y)-M^2 y(1-y)(2\ln[|m_j^2 y+\mu_l^2(1-y)-M^2 y(1-y)|]+1)\right. 
    \nonumber \\
&& \rule{0.5in}{0mm}  \left.
      -(m_j^2 y+\tilde\mu_l^2(1-y)-M^2 y(1-y)(2\ln[|m_j^2 y+\tilde\mu_l^2(1-y)-M^2 y(1-y)|]+1)
           \right]  \nonumber \\
&& \left.+\frac{1}{\mu_l^2}\left[ y(m_j^2+m_{i'}m_{i''})+\frac{2\tilde\mu_l^2(1-y)}{y}\right]
    \ln\left(\left|\frac{m_j^2 y+\tilde\mu_l^2(1-y)-M^2 y(1-y)}
                   {m_j^2 y+\mu_l^2(1-y)-M^2 y(1-y)}\right|\right)\right\}.  \nonumber
\eea
Collecting terms with logarithms, we find that multipliers containing $1/y$ cancel.
The remaining terms containing $1/y$ are not singular in the full expression,
which can be arranged explicitly by adding $\frac{1-\zeta}{\zeta}\frac{1-y}{y}$
to the curly brackets of (\ref{eq:rawnorm}).  This additional piece makes no
contribution to the sum over $l$, because $\sum_l (-1)^l\xi_l=0$.  The
resulting expression for the normalization condition is
\bea                   
1&=&z_0^2-z_1^2+\frac{e^2}{16\pi^2}\sum_{jl}\sum_{i'i''}(-1)^{j+l+i'+i''}\xi_l z_{i'}z_{i''}
  \int dy \\
&& \left\{ \frac{1}{m_j^2 y+\mu_l^2(1-y)-M^2 y(1-y)}
\left[\rule{0mm}{0.3in} 2y(m_j^2-2m_j(m_{i'}+m_{i''})(1-y)+m_{i'}m_{i''}(1-y)^2) 
                      \right.  \right. \nonumber \\
&& \rule{2.5in}{0mm}  \left.
      -\frac{1-\zeta}{\zeta}m_j(m_{i'}+m_{i''})y(1-y)\right] \nonumber \\
&& -2y\ln[|m_j^2 y+\mu_l^2(1-y)-M^2 y(1-y)|] \nonumber \\
&& +\frac{1}{\mu_l^2}\left[\frac{1}{m_j^2 y+\mu_l^2(1-y)-M^2 y(1-y)}
                          -\frac{1}{m_j^2 y+\tilde\mu_l^2(1-y)-M^2 y(1-y)}\right] \nonumber \\
&& \rule{1in}{0mm} \times 
             \left[ m_j^2 m_{i'}m_{i''}y^3
                         +\tilde\mu_l^2 m_j (m_{i'}+m_{i''})y(1-y)\right] \nonumber \\
&& -\left[\frac{1}{m_j^2 y+\mu_l^2(1-y)-M^2 y(1-y)}
                          -\frac{1}{\zeta}\frac{1}{m_j^2 y+\tilde\mu_l^2(1-y)-M^2 y(1-y)}\right] \nonumber \\
&& \rule{1in}{0mm} \times 
             (1-y)\left[m_j^2-M^2(1-y)\right] \nonumber \\
&& \left.+\frac{1}{\mu_l^2}\left[ y(m_j^2+m_{i'}m_{i''})
                 +2[m_j^2-M^2 (1-y)]\right]
    \ln\left(\left|\frac{m_j^2 y+\tilde\mu_l^2(1-y)-M^2 y(1-y)}
                   {m_j^2 y+\mu_l^2(1-y)-M^2 y(1-y)}\right|\right)\right\}.  \nonumber
\eea
In the $m_1\rightarrow\infty$ limit, this becomes
\bea      \label{eq:m1limit}             
1&=&z_0^2+\frac{\alpha}{4\pi}\sum_l(-1)^l\xi_l z_0^2
  \int dy \\
&& \left\{ \frac{1}{m_0^2 y+\mu_l^2(1-y)-M^2 y(1-y)}
\left[\rule{0mm}{0.3in} 2y(m_0^2-4m_0M(1-y)+M^2(1-y)^2) 
                      \right.  \right. \nonumber \\
&& \rule{2.5in}{0mm}  \left.
      -\frac{1-\zeta}{\zeta}2m_0 M y(1-y)\right] \nonumber \\
&& -2y\ln[|m_0^2 y+\mu_l^2(1-y)-M^2 y(1-y)|] \nonumber \\
&& +\frac{1}{\mu_l^2}\left[\frac{1}{m_0^2 y+\mu_l^2(1-y)-M^2 y(1-y)}
                          -\frac{1}{m_0^2 y+\tilde\mu_l^2(1-y)-M^2 y(1-y)}\right] \nonumber \\
&& \rule{1in}{0mm} \times 
             \left[ m_0^2 M^2y^3
                         +2\tilde\mu_l^2 m_0 M y(1-y)\right] \nonumber \\
&& -\left[\frac{1}{m_0^2 y+\mu_l^2(1-y)-M^2 y(1-y)}
                          -\frac{1}{\zeta}\frac{1}{m_0^2 y+\tilde\mu_l^2(1-y)-M^2 y(1-y)}\right] \nonumber \\
&& \rule{1in}{0mm} \times 
             (1-y)\left[m_0^2-M^2(1-y)\right] \nonumber \\
&& \left.+\frac{1}{\mu_l^2}\left[ y(m_0^2+M^2)
                 +2[m_0^2-M^2 (1-y)]\right]
    \ln\left(\left|\frac{m_0^2 y+\tilde\mu_l^2(1-y)-M^2 y(1-y)}
                   {m_0^2 y+\mu_l^2(1-y)-M^2 y(1-y)}\right|\right)\right\}.  \nonumber
\eea
%

%%%%%%%%%%%%%%%%%%%%%%%%%%%%%%%%

\end{document}